\documentclass[twocolumn]{aastex631}
\usepackage{amsmath}
\usepackage{float}
\usepackage{graphicx} 

\defcitealias{carter_benchmark_2024}{Carter \& May et al.}

\begin{document}

\title{Statistical trends in JWST transiting exoplanet atmospheres}
\author[0000-0002-3263-2251]{Guangwei Fu}
\affiliation{Department of Physics and Astronomy, Johns Hopkins University, Baltimore, MD, USA}
\author[0000-0002-7352-7941]{Kevin B. Stevenson}
\affiliation{JHU Applied Physics Laboratory, 11100 Johns Hopkins Rd, Laurel, MD 20723, USA}
\author[0000-0001-6050-7645]{David K. Sing}
\affiliation{Department of Physics and Astronomy, Johns Hopkins University, Baltimore, MD, USA}

\author[0000-0003-1622-1302]{Sagnick Mukherjee}
\affiliation{Department of Astronomy and Astrophysics, University of California, Santa Cruz, CA 95064, USA \\ }
\affiliation{Department of Physics and Astronomy, Johns Hopkins University, Baltimore, MD, USA \\ }

\author[0000-0003-0156-4564]{Luis Welbanks}
\affiliation{School of Earth and Space Exploration, Arizona State University, Tempe, AZ, USA}

\author[0000-0002-5113-8558]{Daniel Thorngren}
\affiliation{Department of Physics and Astronomy, Johns Hopkins University, Baltimore, MD, USA}

\author[0000-0002-8163-4608]{Shang-Min Tsai}
\affiliation{Department of Earth and Planetary Sciences, University of California, Riverside, CA, USA}

\author[0000-0002-8518-9601]{Peter Gao}
\affiliation{Carnegie Science Earth and Planets Laboratory, 5241 Broad Branch Road, NW, Washington, DC 20015, USA}

\author[0000-0003-3667-8633]{Joshua Lothringer}
\affiliation{Space Telescope Science Institute, Baltimore, MD, USA}

\author[0000-0002-9539-4203]{Thomas G. Beatty}
\affiliation{Department of Astronomy, University of Wisconsin-Madison, Madison, WI, USA}

\author[0009-0007-9356-8576]{Cyril Gapp}
\affiliation{Max-Planck-Institut f\"ur Astronomie, K\"onigstuhl 17, D-69117 Heidelberg, Germany}

\author[0000-0001-5442-1300]{Thomas M. Evans-Soma}
\affiliation{School of Information and Physical Sciences, University of Newcastle, Callaghan, NSW, Australia}
\affiliation{Max-Planck-Institut f\"ur Astronomie, K\"onigstuhl 17, D-69117 Heidelberg, Germany}

\author[0000-0002-1199-9759]{Romain Allart}
\affiliation{Institut Trottier de Recherche sur les Exoplan\`etes and D\'epartement de Physique, Universit\'e de Montr\'eal}

\author[0000-0002-8573-805X]{Stefan Pelletier}
\affiliation{Observatoire astronomique de l’Universit\'e de Gen\`eve, 51 chemin Pegasi 1290 Versoix, Switzerland}

\author[0000-0001-5729-6576]{Pa Chia Thao}
\affiliation{Department of Physics and Astronomy, The University of North Carolina at Chapel Hill, Chapel Hill, NC 27599, USA}

\author[0000-0003-3654-1602]{Andrew W. Mann}
\affiliation{Department of Physics and Astronomy, The University of North Carolina at Chapel Hill, Chapel Hill, NC 27599, USA}

\begin{abstract}

Our brains are hardwired for pattern recognition as correlations are useful for predicting and understanding nature. As more exoplanet atmospheres are being characterized with JWST, we are starting to unveil their properties on a population level. Here we present a framework for comparing exoplanet transmission spectroscopy from 3 to 5$\mu$m with four bands: L (2.9 - 3.7$\mu$m), SO$_2$ (3.95 - 4.1$\mu$m), CO$_2$ (4.25 - 4.4$\mu$m) and CO (4.5 - 4.9$\mu$m). Together, the four bands cover the major carbon, oxygen, nitrogen, and sulfur-bearing molecules including H$_2$O, CH$_4$, NH$_3$, H$_2$S, SO$_2$, CO$_2$, and CO. Among the eight high-precision gas giant exoplanet planet spectra we collected, we found strong correlations between the SO$_2$-L index and planet mass (r=-0.41$\pm$0.09) and temperature (r=-0.64$\pm$0.08), indicating SO$_2$ preferably exists (SO$_2$-L$>$-0.5) among low mass ($\sim<$0.3M$_J$) and cooler ($\sim<$1200K) targets. We also observe strong temperature dependency for both CO$_2$-L and CO-L indices. Under equilibrium chemistry and isothermal thermal structure assumptions, we find that the planet sample favors super-solar metallicity and low C/O ratio ($<$0.7). In addition, the presence of a mass-metallicity correlation is favored over uniform metallicity with the eight planets. We further introduce the SO$_2$-L versus CO$_2$-L diagram alike the color-magnitude diagram for stars and brown dwarfs. All reported trends here will be testable and be further quantified with existing and future JWST observations within the next few years. 

\end{abstract}

\nopagebreak
\section{Introduction}

Statistical trends offer us insights into hidden patterns in nature. Finding and understanding correlations in data has been a powerful tool in science. Indeed, it has been used across disciplines from studying how mammal metabolic rates change with their body mass \citep{hennemann_relationship_1983} to the effect of deforestation on biodiversity \citep{liang_positive_2016}. The foundation of modern astronomy started in the Hertzsprung-Russell diagram, a correlation between colors of stars. The Tully-Fisher correlation of galaxies gave us a new method of determining cosmic distance \citep{tully_new_1977}. The relationship between black hole mass and the velocity dispersion of its host galaxy tells us how they evolve together in time \citep{kormendy_coevolution_2013}. The color sequence within brown dwarfs \citep{geballe_toward_2002, suarez_ultracool_2022} shows how they cool over time and experience chemical transitions within the atmospheres.

Many major advancements in the study of exoplanets have also been coming from statistical studies. The Kepler planet sample showed that small planets consist of two separate populations \citep{fulton_california-kepler_2017} which motivated further studies of how atmospheric loss shapes small short-period planet demographics \citep{owen_evaporation_2017}. The correlation between planet occurrence rates and host star metallicity provides insights into different planet formation scenarios \citep{mann_they_2012, reffert_precise_2015, lu_increase_2020}. 

\begin{figure*}[htp]
    \centering
    \includegraphics[width=0.8\textwidth]{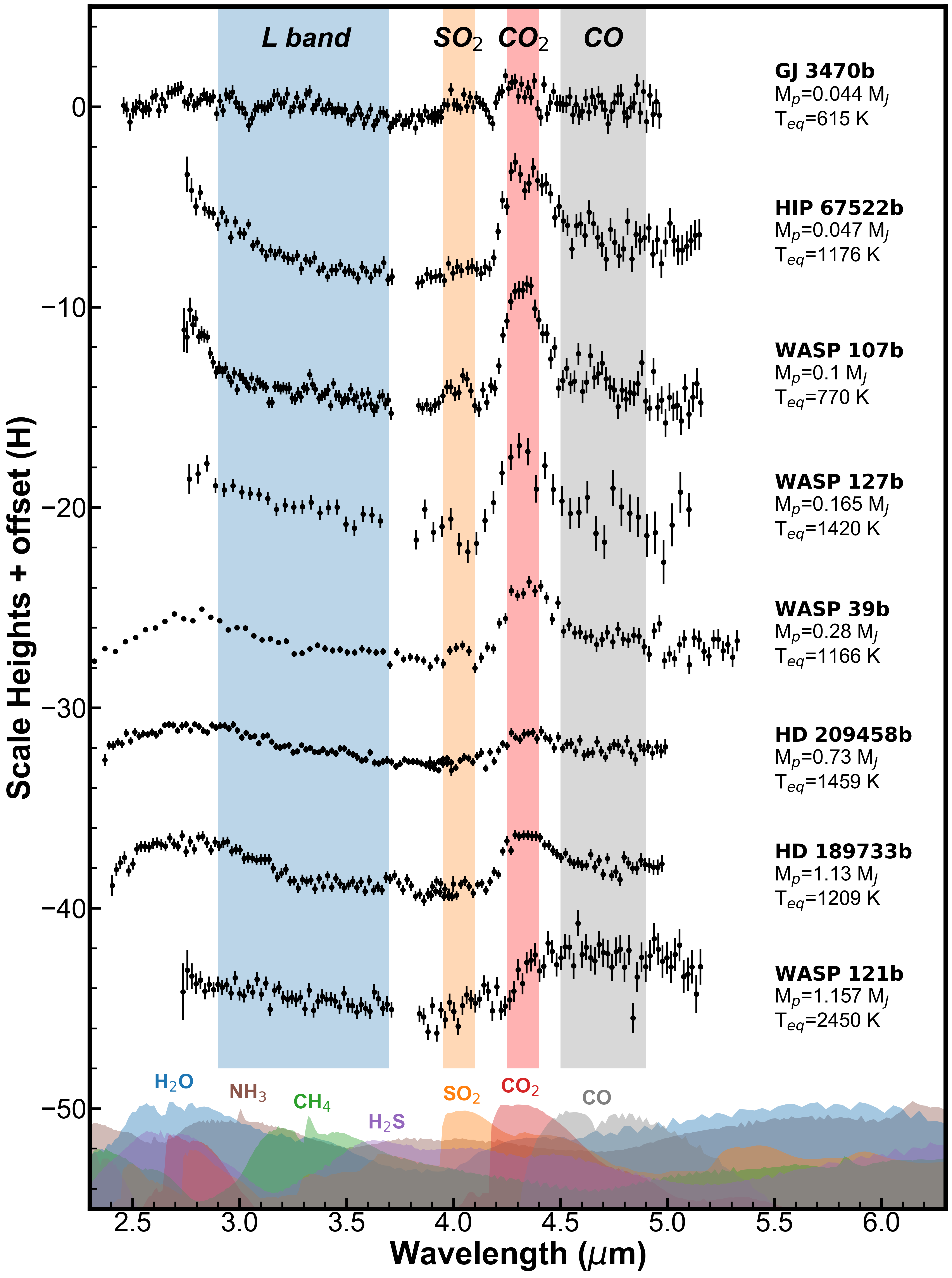}
    \caption{Transmission spectra included in this study. All spectra are normalized by their respective atmospheric scale heights (H) and plotted with a vertical offset. All spectra cover the 2.7 to 5 $\mu$m wavelength range. We picked four bands: L (2.9 - 3.7$\mu$m), SO$_2$ (3.95 - 4.1$\mu$m), CO$_2$ (4.25 - 4.4$\mu$m) and CO (4.5 - 4.9$\mu$m) which are color shaded with blue, orange, red and grey respectively. These four bands cover major oxygen, carbon, and sulfur-bearing molecules such as H$_2$O, CH$_4$, NH$_3$, H$_2$S, SO$_2$, CO$_2$ and CO.}
\label{fig1}   
\end{figure*} 

We are now starting to see tentative statistical trends in exoplanet atmospheric measurements, which have grown drastically in volume in the past decade. Early studies focused on HST data from optical to the near-infrared. In the optical, we have seen varying cloud properties as shown by the optical scattering slope \citep{sing_continuum_2016, pinhas_signatures_2017, fisher_retrieval_2018, wakeford_exoplanet_2019} with increasing transit depth at shorter wavelength. In the near-infrared, \cite{stevenson_quantifying_2016} first introduced the method of calculating in-and-out of 1.4 $\mu$m water band difference in scale heights and using it as an index to compare different exoplanet atmospheres. The correlation between the water band index and temperature \citep{stevenson_quantifying_2016, fu_statistical_2017, crossfield_trends_2017} indicates temperature-dependent aerosols properties \citep{gao_aerosol_2020, brande_clouds_2023}. The infrared trends from Spitzer on relative eclipse depth of the two channels at 3.6 and 4.5 $\mu$m versus temperature \citep{deming_emergent_2023, wallack_trends_2021, mansfield_unique_2021, baxter_evidence_2021} could suggest the onset of disequilibrium atmospheric processes and changing thermal structures.

The launch of JWST has given us an order-of-magnitude improvement in photometric precision and spectroscopic access into the infrared. JWST observations so far have already brought new discoveries such as the detection of SO$_2$ \citep{rustamkulov_early_2023, tsai_photochemically_2023, dyrek_so2_2023, alderson_early_2023, powell_sulfur_2024} and spectrally resolved CO$_2$ \citep{the_jwst_transiting_exoplanet_community_early_release_science_team_identification_2022, xue_jwst_2023, fu_hydrogen_2024}. The molecules SO$_2$ and CO$_2$ are both predicted to be sensitive atmospheric metallicity tracers \citep{tsai_photochemically_2023, moses_chemical_2013}, and metallicity has been proposed as a key parameter to test different planet formation scenarios \citep{mordasini_imprint_2016}. With many transiting exoplanets that have already been observed with JWST, there has yet to be a population-level study of their atmospheres in the JWST infrared wavelength ranges. Here we collected eight JWST transit exoplanet spectra across an order of magnitude in mass and span over 1000K in temperature. By correlating their atmospheric features with parameters such as planet mass, equilibrium temperature, host star type, etc, we aim to gain new physical insights into atmospheric chemistry under different conditions and scaffold a color sequence for transiting exoplanets.

\section{Methods}

\subsection{Sample selection}

These eight planets (Table \ref{tab:table1}, \ref{tab:table2}) were selected among all available JWST spectra based on three criteria: (1) Transit spectrum coverage from 3-5 $\mu m$, (2) High precision spectrum where SO$_2$ and CO$_2$ bands can be robustly measured, and (3) Planet mass $>$ 0.03 M$_J$. These three criteria ensure a uniform and robust population study among transiting hydrogen-dominated giant planet atmospheres.

\subsection{WASP-127b G395H transit spectrum}

We observed a transit of WASP-127b using JWST NIRSpec G395H (GO 2437 PI: Stefan Pelletier) on May 8th, 2023. Due to an issue in the observation setup, the star is not fully placed in the slit and $\sim$10$\%$ of the total flux is captured which leads to a lower-than-expected spectral precision (more details in Allart et al. in prep). However, the excellent JWST pointing stability ($\sim$0.01 pixels) throughout the observation means the effects of slit loss are minor. We performed the data reduction following the steps described by \cite{rustamkulov_early_2023} and \cite{sing_warm_2024}. The orbital parameters (a/R$_s$ and inclination) were fixed to values from \cite{seidel_hot_2020}. The reduced spectrum is shown in Figure \ref{fig1} and used for the analysis in this paper.

\subsection{Spectra normalization}
Our eight JWST transit spectra (available at https://doi.org/10.5281/zenodo.13366851) have wavelength coverage between $\sim$2.8 and 5$\mu$m (Table \ref{tab:table1}). Each spectrum is normalized by its atmospheric scale height, $H$, which is calculated via equation H = kT/$\mu$g with k being the Boltzmann constant, T being the temperature, $\mu$ being the mean molecular weight, and g being the surface gravity. For T, we used the equilibrium temperature (Table \ref{tab:table1}) for all planets under the assumption that it is a close approximation of the atmospheric temperature at the pressure levels ($\sim$1-10 mbar) that transmission spectroscopy probes. For $\mu$, we adopted 2.3 amu for all planets which may not be true for planets with significantly different atmospheric composition and metal enrichment levels. However, since the variation of atmospheric metallicity with other parameters, such as planet mass and orbital period, is a key statistical correlation of interest, the exact value of $\mu$ used is not important as long as it is uniformly applied to all planets within the data and corresponding model. For g, we used values calculated from the reported planet's mass and radius in the previous literature. The sequence of all spectra normalized by their scale heights is plotted in Figure \ref{fig1} with a constant vertical offset.

\begin{table*}[htp]
\centering
\begin{tabular}{ccccc}
\hline
\hline
Planet	&	JWST mode	&	Spectrum   &   Program ID  &	Planet parameter Ref.	\\
\hline
GJ 3470b	&	NIRCam F322W2 + F444W	&	\cite{beatty_sulfur_2024}	&	GTO 1185	&	\cite{awiphan_transit_2016}	\\
HIP 67522b	&	NIRSpec G395H	&	Thao et al. (Submitted)	&	GO 2498	&	Thao et al. (Submitted)	\\
WASP 107b	&	NIRSpec G395H	&	\cite{sing_warm_2024}	&	GTO 1201	&	\cite{anderson_discoveries_2017}	\\
WASP 127b	&	NIRSpec G395H	&	This work	&	GO 2437	&	\cite{seidel_hot_2020}	\\
WASP 39b	&	NIRSpec PRISM	&	\citetalias{carter_benchmark_2024}	&	ERS 1366	&	\cite{mancini_gaps_2018}	\\
HD 209458b	&	NIRCam F322W2 + F444W	&	\cite{xue_jwst_2023}	&	GTO 1274	&	\cite{stassun_accurate_2017}	\\
HD 189733b	&	NIRCam F322W2 + F444W	&	\cite{fu_hydrogen_2024}	&	GO 1633	&	\cite{stassun_accurate_2017}	\\
WASP 121b	&	NIRSpec G395H	&	Gapp et al. (Submitted)	&	GO 1729	&	\cite{bourrier_hot_2020}	\\
\hline
\end{tabular}
\caption{Planet included in Figure \ref{fig1} with their respective JWST observing mode and references.}\label{tab:table1}
\end{table*}

\begin{figure*}[htp]
    \centering
    \includegraphics[width=0.8\textwidth]{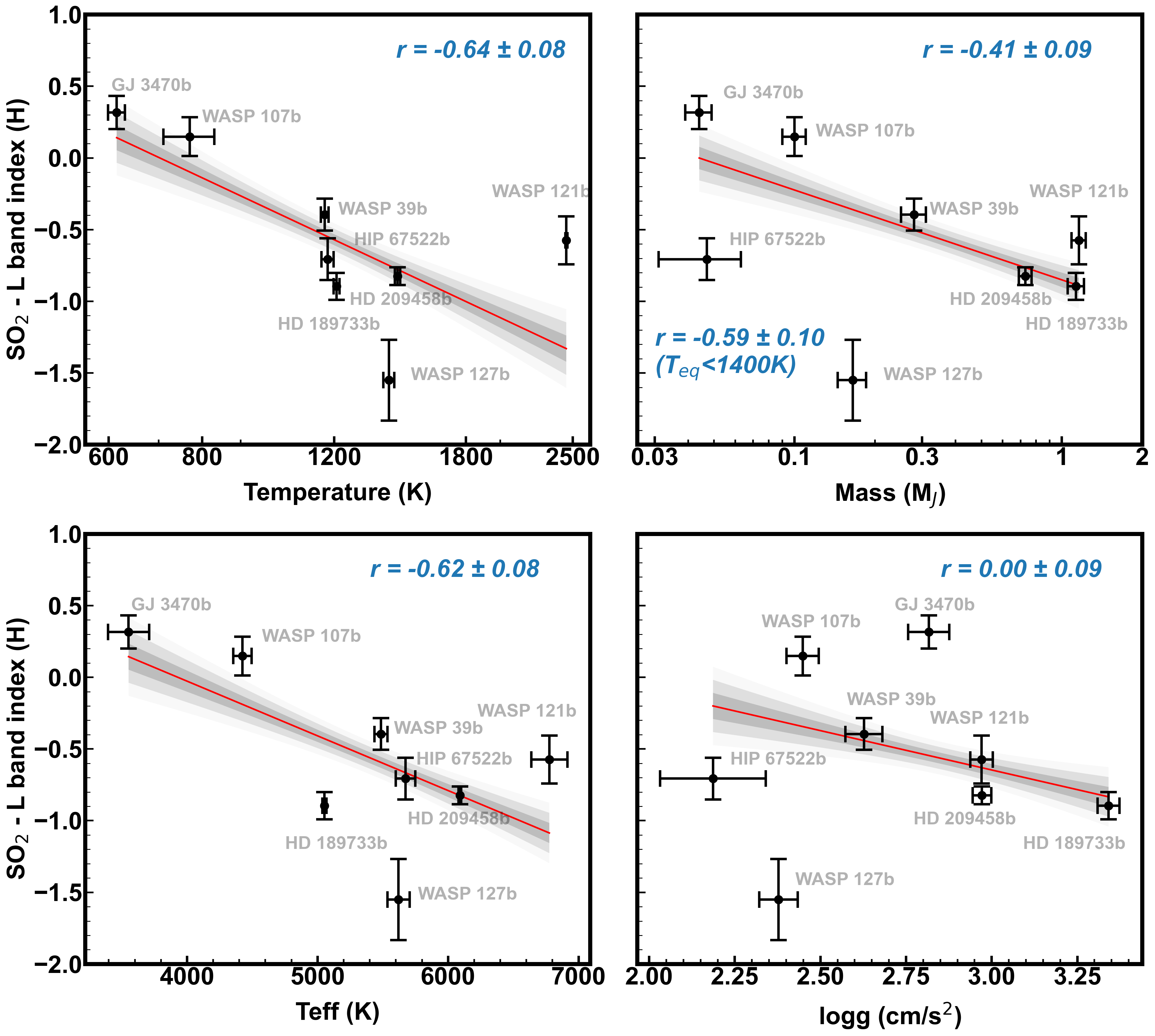}
    \caption{SO$_2$-L versus equilibrium temperature (upper left), planet mass (upper right), host star effective temperature (lower left), and planet surface gravity (lower right). The red lines indicate the best-fit trend line and the grey shaded regions represent one to three sigma uncertainties.}
    \label{fig2}   
\end{figure*}

\begin{figure}[t]
    \centering
    \includegraphics[width=0.45\textwidth]{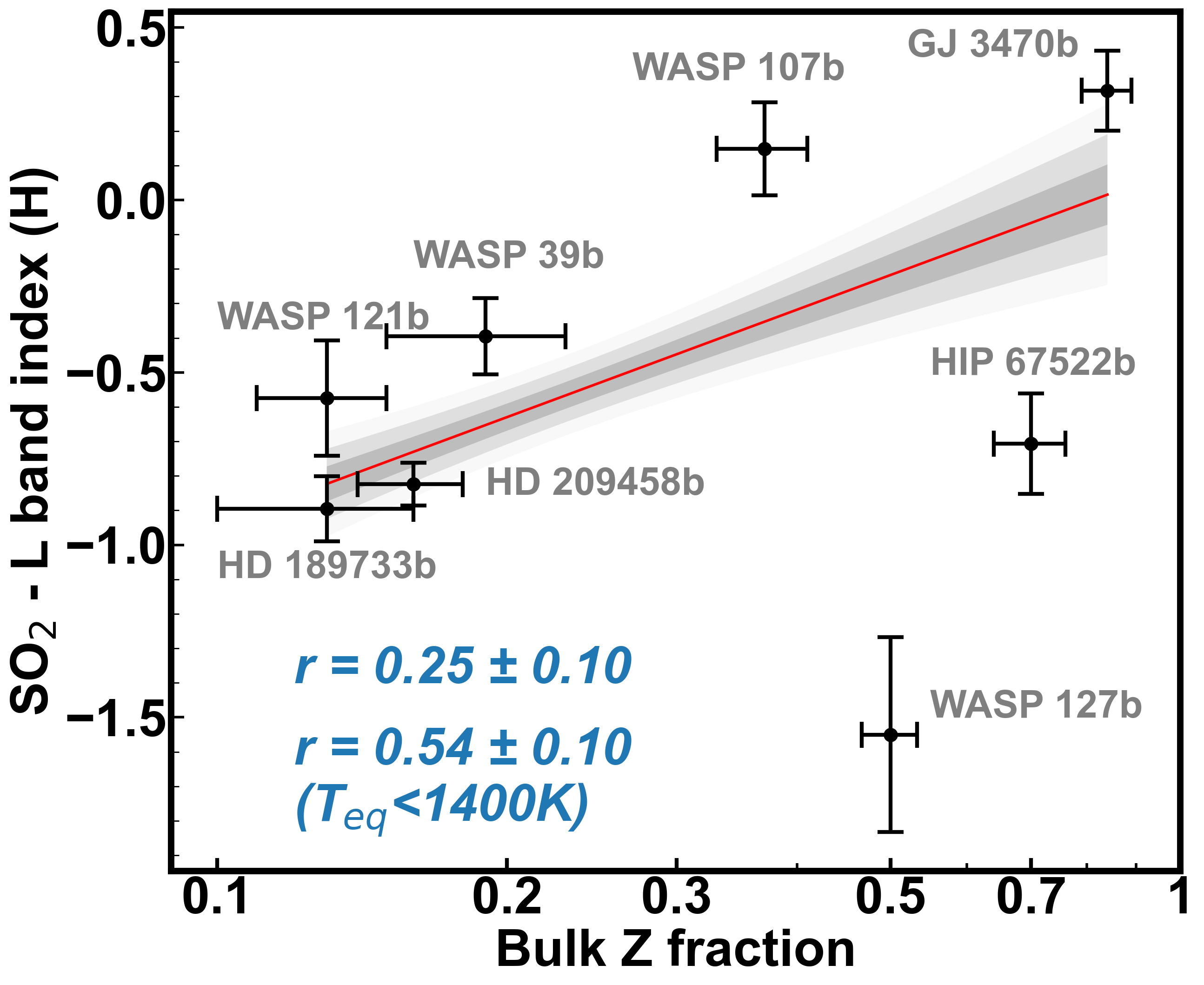}
    \caption{SO$_2$-L versus planet bulk Z fraction \citep{thorngren_massmetallicity_2016, thorngren_connecting_2019}. The positive correlation between the two suggests that SO$_2$-L index traces the metallicity enrichment levels in the planet.}
\label{figZ}   
\end{figure}

\begin{figure*}[t]
    \centering
    \includegraphics[width=1\textwidth]{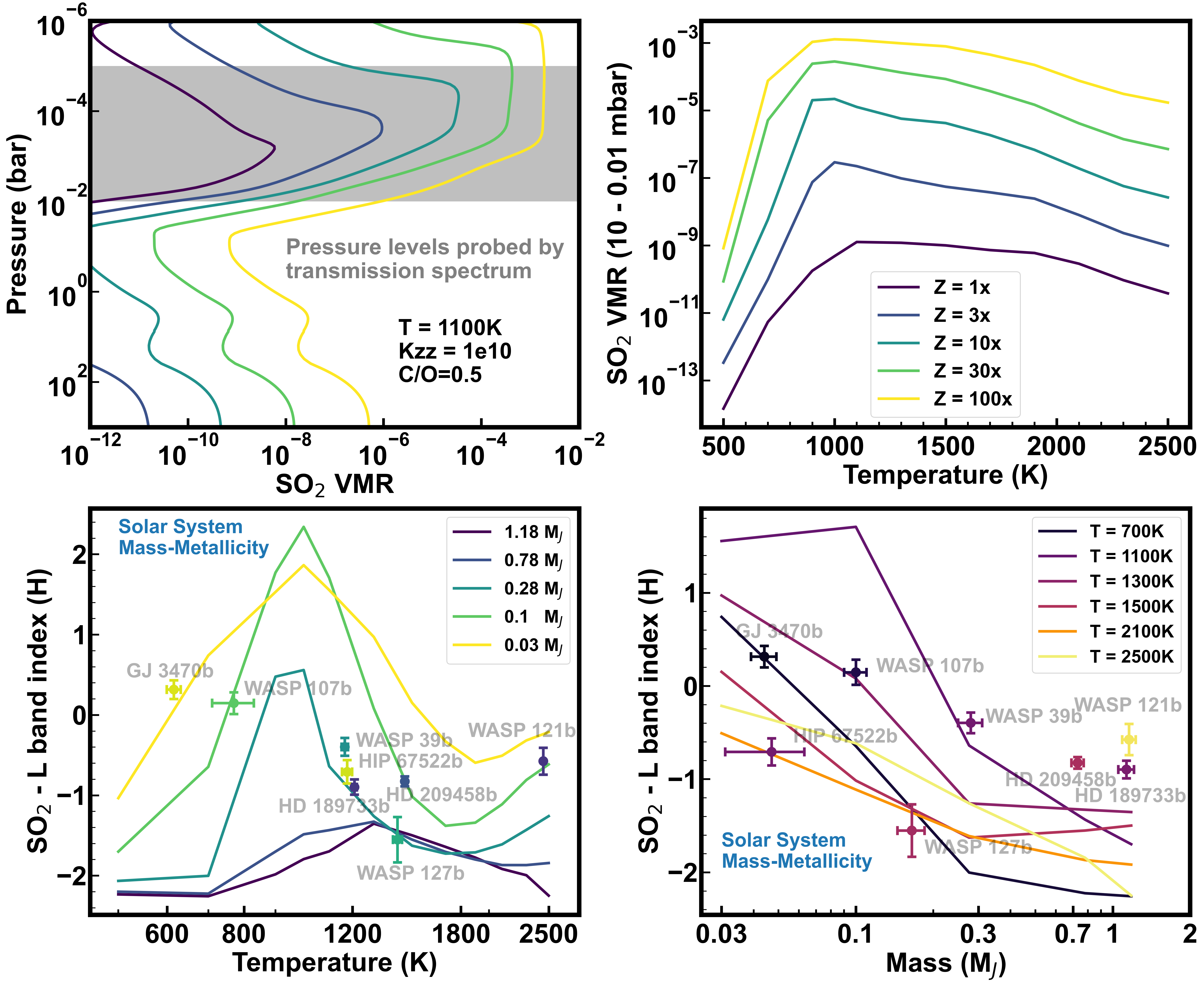}
    \caption{Volume mixing ratio of SO$_2$ in the atmosphere as a function of metallicity for the model grid at 1100K (Top left). Average SO$_2$ VMR across 10 to 0.01 mbar pressure levels versus temperature for five metallicity values (Top right). The model grid-predicted SO$_2$-L values versus temperature for the five metallicity values are shown in the bottom left, with the metallicities converged to planet masses (shown in color) assuming the solar system mass-metallicity relation. The bottom right shows the SO$_2$-L model values versus planet mass at constant temperatures, again with the model metallicities converted to mass via the solar system mass-metallicity relation.}
    
\label{SO2_models}   
\end{figure*}

\begin{figure*}[t]
    \centering
    \includegraphics[width=0.8\textwidth]{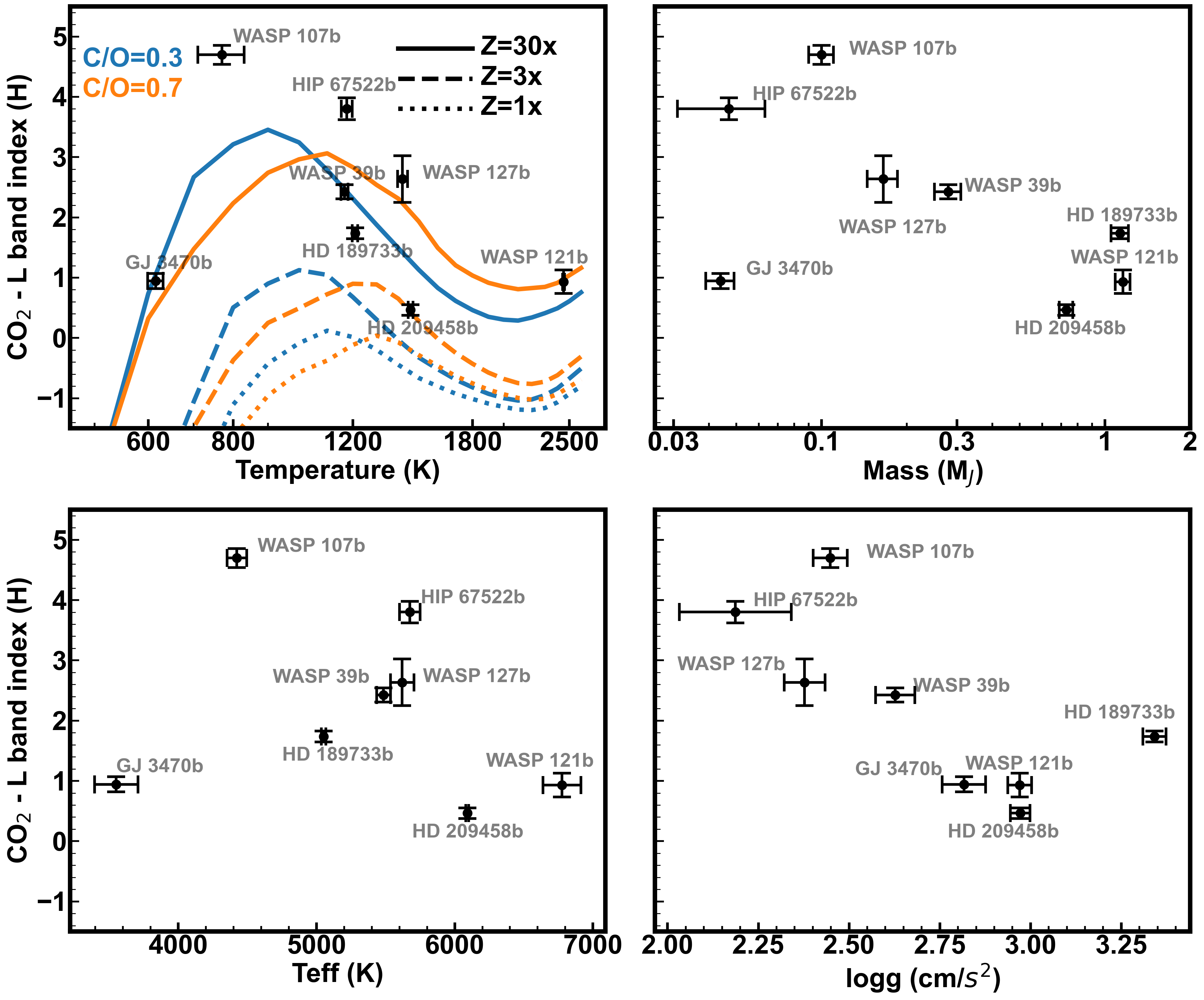}
    \caption{CO$_2$-L versus equilibrium temperature (upper left), planet mass (upper right), host star effective temperature (lower left), and planet surface gravity (lower right). There are no clear linear trends, which is expected as CO$_2$ is sensitive to temperature non-monotonically. We overplotted two sets of forward \texttt{PLATON} models with different Z and C/O in the top left panel.}
\label{fig3}   
\end{figure*}

\begin{figure*}[t]
    \centering
    \includegraphics[width=0.8\textwidth]{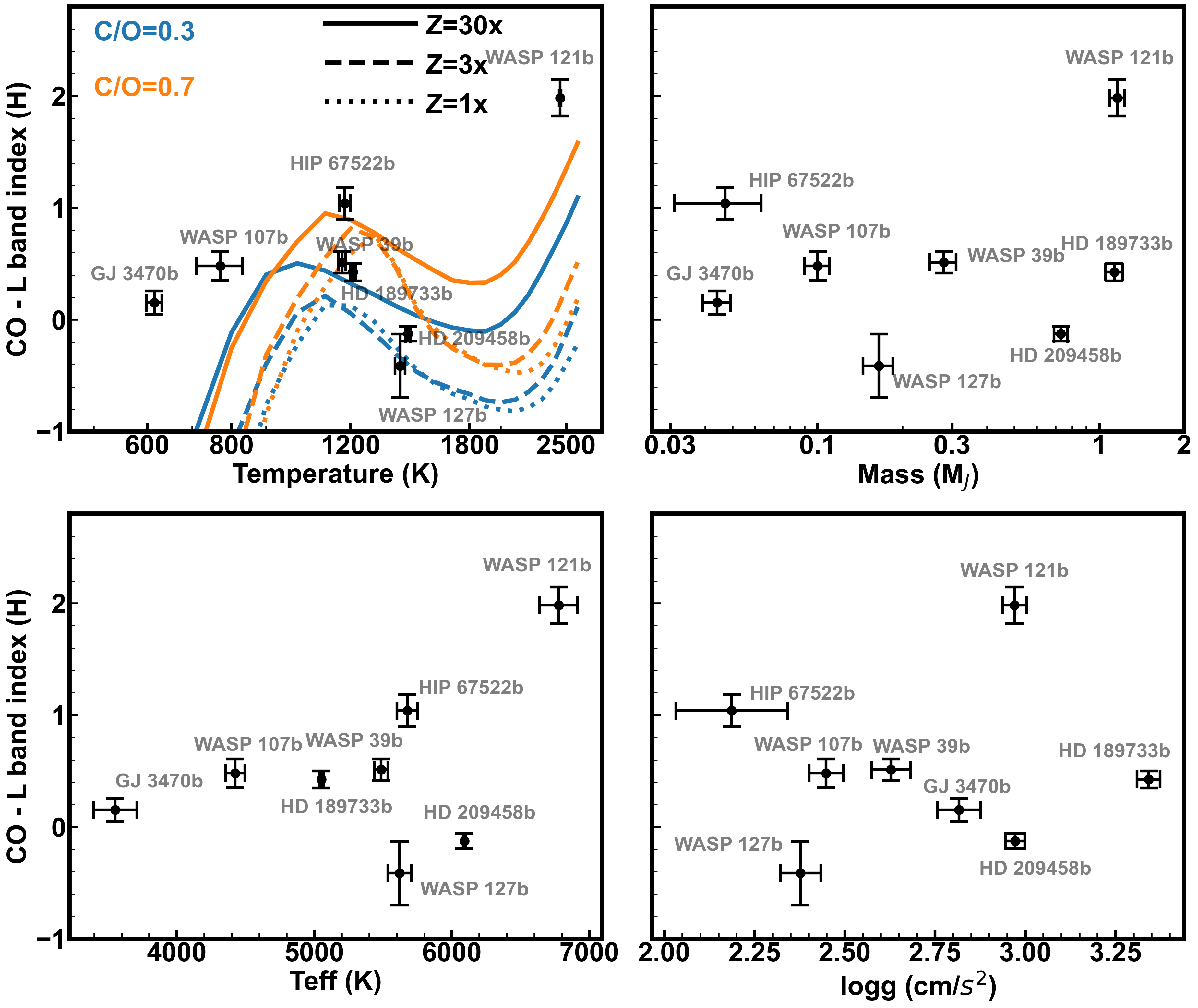}
    \caption{CO-L versus equilibrium temperature (upper left), planet mass (upper right), host star effective temperature (lower left), and planet surface gravity (lower right). As temperature increases, CO becomes the major carbon-bearing molecule in the atmosphere. We overplotted two sets of forward \texttt{PLATON} models with different Z and C/O in panel two.}
\label{fig3_CO}   
\end{figure*}

\begin{figure*}[t]
    \centering
    \includegraphics[width=1\textwidth]{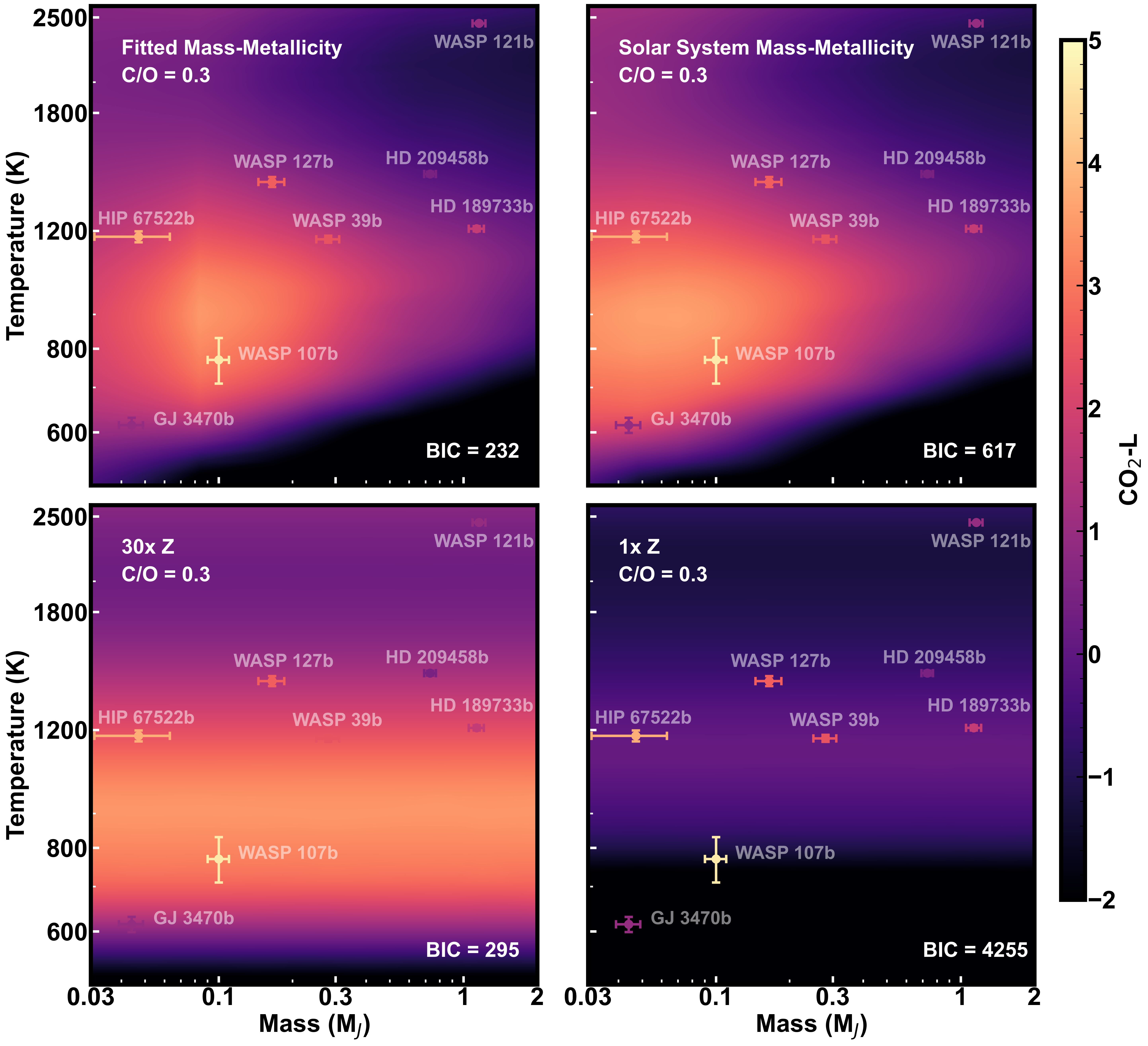}
    \caption{Planet mass versus equilibrium temperature with each point color-coded representing their CO$_2$-L index value. The background color gradients are model predictions from the PLATON generic grid under Solar System mass-metallicity trend or uniform metallicity assumptions.}
\label{fig_SS}   
\end{figure*}

\begin{figure*}[t]
    \centering
    \includegraphics[width=1\textwidth]{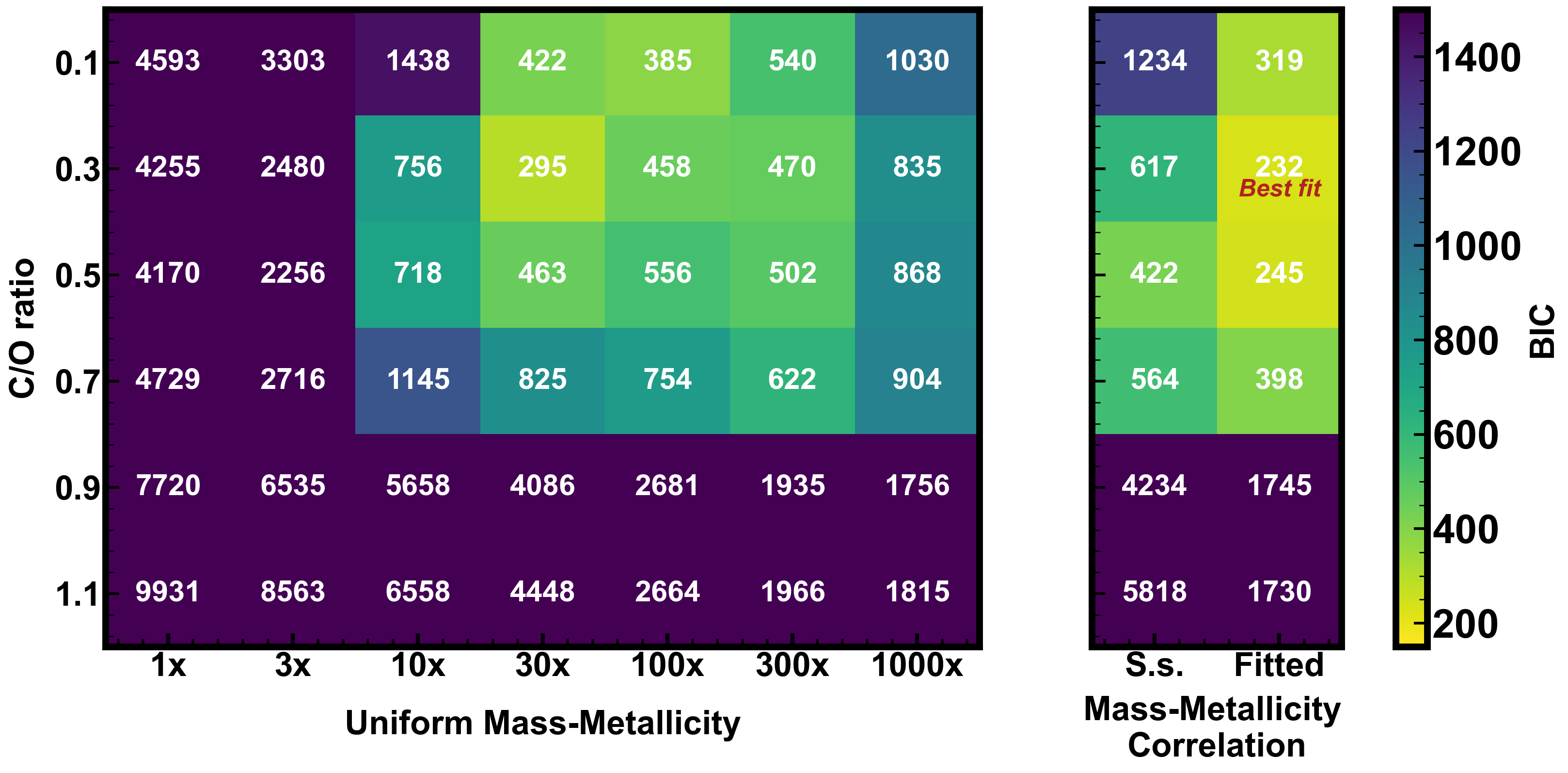}
    \caption{The scale height normalized equilibrium chemistry and isothermal \texttt{PLATON} forward models are compared to the CO$_2$-L index from the sample. The C/O ratio is fixed for the entire sample for each row. The left panel assumes the same metallicity for all planets at each column. The right panel assumes solar-system (S.s) mass-metallicity correlation and freely fitted correlation. The BIC values are shown for each grid point. At the population level, low metallicity ($<$ 10xZ) and high C/O ($>$0.9) are strongly disfavored. The presence of a mass-metallicity correlation is favored.}
\label{fig_BIC}
\end{figure*}

\begin{figure*}[t]
    \centering
    \includegraphics[width=1\textwidth]{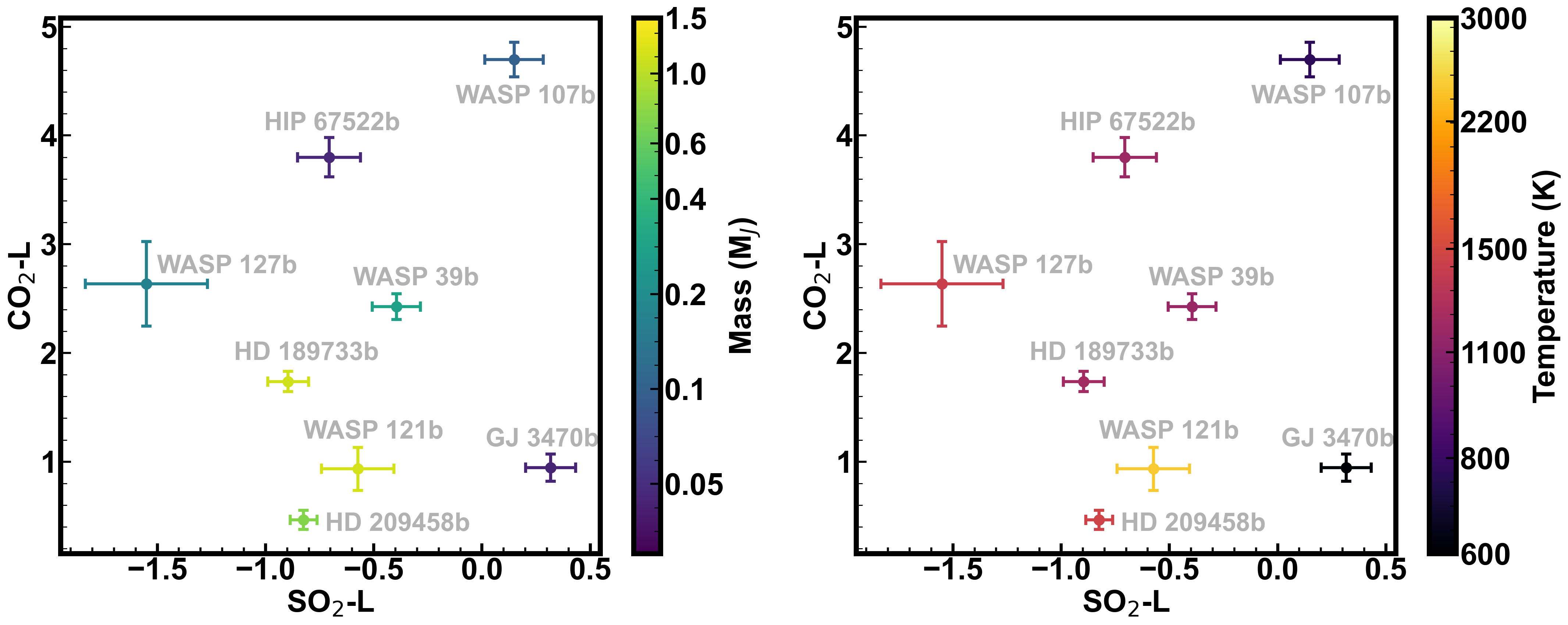}
    \caption{The SO$_2$-L versus CO$_2$-L diagrams with color scales for planet mass (left) and temperature (right). No clear patterns have emerged from the limited sample. We present a framework for population-level exoplanet atmosphere characterization as future JWST transmission spectra fill in this diagram.}
\label{fig4}   
\end{figure*}

\begin{figure*}[t]
    \centering
    \includegraphics[width=1\textwidth]{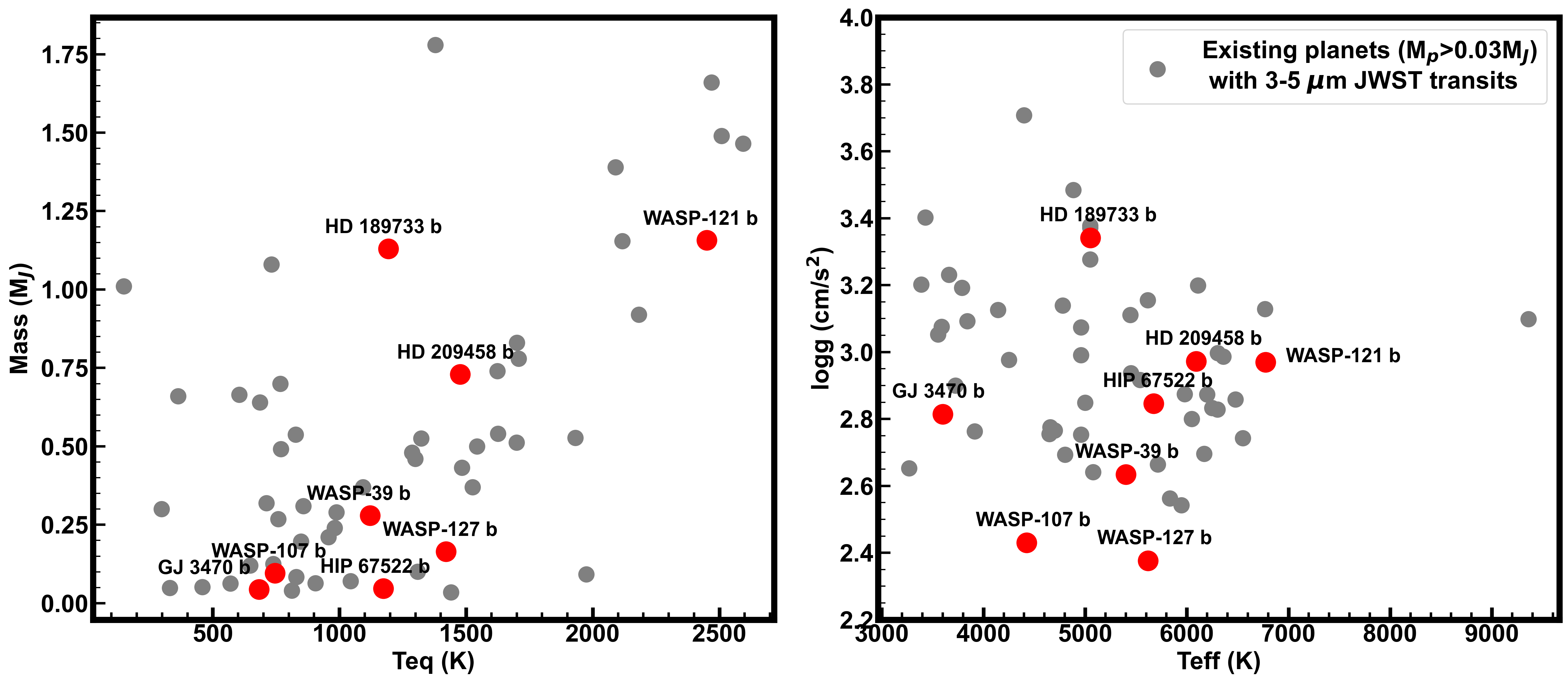}
    \caption{The eight planets included in this study are correlated in their respective physical parameters such as mass and temperature. However, there are 56 planets in the JWST cycles 1, 2, and 3 programs that will have transmission spectra covering the 3 to 5$\mu$m wavelength range. As more planet transmission spectra are analyzed and added to this diagram, we will resolve the degeneracies and test if these trends and tracks between L, SO$_2$, CO$_2$, and CO bands remain statistically significant.}
\label{fig5}   
\end{figure*}

\subsection{SO$_2$, CO$_2$, CO and L}
To compare these normalized spectra, we focus on four photometric bands: L (2.9 - 3.7$\mu$m), SO$_2$ (3.95 - 4.1$\mu$m), CO$_2$ (4.25 - 4.4$\mu$m) and CO (4.5 - 4.9$\mu$m). The goal of selecting these four bands is to use them to best capture the changes in the transit spectra from varying atmospheric compositions as the underlying physical properties (Temperature, Mass, etc…) of the planet change. The L band covers H$_2$O, CH$_4$, NH$_3$, and H$_2$S features. The wide L band width is chosen to best match the broad opacity shapes of molecules in this wavelength range. The SO$_2$, CO$_2$, and CO bands are centered on SO$_2$, CO$_2$, and CO band heads respectively. These four bands cover the major carbon, oxygen, and sulfur-bearing molecules present within this wavelength range (Figure \ref{fig1}). H$_2$O, CO$_2$ and SO$_2$ are sensitive to overall metallicity \citep{moses_disequilibrium_2011, tsai_photochemically_2023}, CO and CH$_4$ responds to C/O and vertical mixing changes \citep{moses_disequilibrium_2011}, SO$_2$ and NH$_3$ trace photochemistry \citep{tsai_comparative_2021} and H$_2$S reflects equilibrium sulfur abundance. Their relative differences indicate the relative strength of corresponding bands and are informative in uncovering the population-level patterns and trends within exoplanet atmospheres. 

Sulfur dioxide is one of the major discoveries from transiting exoplanet atmospheric characterization with JWST so far. It was first seen in the Early Release Science program \citep{rustamkulov_early_2023, tsai_photochemically_2023, alderson_early_2023, powell_sulfur_2024} on WASP-39b, a hot and inflated Saturn-mass exoplanet. It was later also detected on the low-mass giant WASP-107b \citep{dyrek_so2_2023}. However, similar SO$_2$ features were not seen on other hot Jupiters \citep{fu_hydrogen_2024, xue_jwst_2023}. We know SO$_2$ is formed photochemically via the oxidation of sulfur radicals generated from the destruction of hydrogen sulfide, but the physical conditions needed to form SO$_2$ are still poorly constrained from observations. Previous studies have predicted that SO$_2$ abundance could be sensitive to atmospheric metallicity, temperature, and stellar UV flux \citep{zahnle_atmospheric_2009, hobbs_sulfur_2021, tsai_comparative_2021, polman_h_2023, tsai_photochemically_2023}. 

To search for any empirical correlations between SO$_2$ feature sizes and physical parameters, we compute the average transit depth in scale heights within the SO$_2$ band covering 3.95 and 4.1 $\mu$m for each planet. Since transmission spectroscopy is a relative measurement of planet-to-stellar radius ratio at different wavelengths and not an absolute flux measurement, the SO$_2$ feature scale height value by itself is not informative and we need to compare it to another band. Here we compare it to the L band by subtracting the L band (2.9 to 3.7 $\mu$m) value from it to obtain the relative SO$_2$ - L index. The uncertainties are added in quadrature. We correlate the SO$_2$ - L index with planet mass, temperature, surface gravity, and host star effective temperature (Figure \ref{fig2}). These parameters were selected based on theoretical predictions that suggest planet mass determines bulk composition, temperature governs chemical processes, surface gravity affects atmospheric dynamics, and effective temperature traces the stellar UV flux. 


Carbon dioxide has been another major discovery from JWST with its prominent 4.25 to 4.4 $\mu$m absorption feature first spectrally resolved by \citet{the_jwst_transiting_exoplanet_community_early_release_science_team_identification_2022}. Between $\sim$800K to 2000K, CO$_2$ is expected to be relatively insensitive to disequilibrium chemistry in H$_2$-dominated atmospheres \citep{moses_disequilibrium_2011} and a robust tracer for metallicity enhancement at a given temperature \citep{lodders_atmospheric_2002, moses_disequilibrium_2011}. Similar to SO$_2$ - L, we compute the mean transit depth in scale heights within the CO$_2$ band and subtract the L band value from it to obtain the CO$_2$ - L index. We also correlate it with planet mass, temperature, surface gravity, and host star effective temperature (Figure \ref{fig3}).

Carbon monoxide is a key molecule for hot to ultra-hot Jupiters since most of the carbon in high-temperature H$_2$-dominated atmospheres ($>$1000K) is expected to be in CO \citep{moses_chemical_2013}. To obtain the CO - L index, we average the transit depth in scale heights within the CO band covering 4.5 and 4.9 $\mu$m and subtract the L band value.

\section{Discussion}

\subsection{SO$_2$-L trends}

We found strong negative statistical correlations between SO$_2$-L and planet temperature and host star effective temperature (r$<$-0.5). The correlation is moderate (-0.41$\pm$0.09) with planet mass, while no correlation was found between SO$_2$-L and surface gravity. We also calculated the planet bulk metallicity (Z) fraction \citep{thorngren_connecting_2019} which represents the upper limit of atmospheric metallicity if the entire planet is well mixed (Figure \ref{figZ}). To compute the bulk metallicities of the planets in our sample, we use the approach of \cite{thorngren_connecting_2019}, solving the equations of hydrostatic equilibrium, mass conservation, and the material equations of state (EOS).  We use the \cite{chabrier_new_2019} EOS for hydrogen and helium, and the ANEOS EOS \citep{thompson_aneos_1990} for the metals, which we assume to be a 50-50 mixture of water and rock.  Thermal evolution is conducted by integrating the rate at which heat is released in the interior using the \cite{fortney_planetary_2007} atmosphere models.  For planets above 1000 K, the hot Jupiter heating is set according to the \cite{thorngren_bayesian_2018} flux-heating relationship. WASP-107 b is an exception; because it appears to be tidally heated (e.g. \cite{millholland_tidal_2020}), we adopt the bulk metallicity of \cite{sing_warm_2024}, which had been calculated in the same way as our other planets except that its adiabat entropy was set by the observed intrinsic temperature rather than normal thermal evolution.

SO$_2$-L shows a weak correlation (r=0.25$\pm$0.1) with bulk Z fraction but the correlation becomes stronger (r=0.54$\pm$0.1) for planets cooler than WASP-127b ($\sim$1400K) where SO$_2$ feature is expected to diminish with higher temperature \citep{tsai_photochemically_2023} (Figure \ref{SO2_models}).

\subsubsection{SO$_2$-L versus temperature and mass}

To further interpret the observed trends and better explain the scatter, we ran a set of radiative-convective equilibrium forward models using the \texttt{PICASO} \citep{batalha_exoplanet_2019, mukherjee_picaso_2023} climate model coupled with the  \texttt{photochem} \citep{wogan_origin--life_2023} 1D chemical kinetics model for photochemistry. We assumed a constant planet mass of 0.42M$_{\rm jup}$ and 1 Jupiter radius, giving a gravity of 10 m/s$^2$. The atmosphere was assumed to be cloud-free with solar C/O and a vertically constant K$_{zz}$=10$^{10}$ cm$^2$/s. The T$_{\rm int}$ was assumed to be 200 K. The model grid covers metallicity from 1x to 100x solar and equilibrium temperatures from 500K to 2500K. All model transmission spectra are then normalized with the scale height using the gravity of 10 m/s$^2$ and a mean molecular weight of 2.3 amu. This normalization step allows for model comparison to planets with different masses.

The SO$_2$ abundance has been predicted to vary with metallicity and temperature \citep{polman_h_2023, tsai_photochemically_2023}. Here we first show the vertical VMR of SO$_2$ as predicted by our grid (Top left panel of Figure \ref{SO2_models}) at 1100K. The grey region represents the pressure levels (10 - 0.01 mbar) typically probed by transmission spectroscopy. We then average the SO$_2$ VMR within that region and plot them against temperature and metallicity (Top right panel of Figure \ref{SO2_models}). At a given temperature, SO$_2$ abundance is expected to increase with metallicity. At a given metallicity, SO$_2$ abundance is expected to peak around 1000K. Next, we show how these model predictions compare to the observed SO$_2$-L trends with temperature and mass. We calculated the SO$_2$-L from the models and overplotted the measured data values (Bottom left panel of Figure \ref{SO2_models}). All data points are color-coded with their mass and model lines are converted to the same color scale assuming the solar system CH$_4$ abundance mass metallicity correlation log(CH$_4$/H) = -1.11 log (M/M$_J$) + 0.38 \citep{atreya_origin_2016}. In other words, all model lines would have the same color under a uniform mass-metallicity relation. The cooler planets GJ 3470b and WASP 107b have high SO$_2$-L values and drive the temperature empirical trend. High metallicity is needed to fit these two points. The hotter planets with lower SO$_2$-L values scatter around -1 where multiple model predictions with low Z($<$10x)/high mass($<$0.28M$_J$) converge due to the lack of SO$_2$ feature under unfavorable high temperature and low Z environments. This shows that the empirical SO$_2$-L versus temperature trend is due to a combination of changing temperature and metallicity/mass in the sample. The data points have excess scatter relative to colored lines, especially around planets with no SO$_2$ feature. This is likely due to the presence of clouds in the observed spectra. Since the models are cloud-free, if the SO$_2$ feature is not present, the SO$_2$-L will be decreased relative to model prediction as the continuum is raised around the 4 $\mu$m region. We also show the SO$_2$-L versus planet mass trend with models and data points color-coded with their respective temperatures (Bottom right panel of Figure \ref{SO2_models}).

\subsubsection{SO$_2$-L versus T$_{eff}$ and logg}

Host star temperature shows significant correlations with SO$_2$-L index while surface gravity does not. SO$_2$ is a photochemical product, and thus the effect of stellar radiation especially the ratio of NUV and FUV flux could be important to its production and destruction. Host star temperature traces the stellar SED to the first order. We observe a correlation between SO$_2$-L and T$_{eff}$ with increasing SO$_2$-L values with decreasing T$_{eff}$. This could indicate that K star SEDs drive higher SO$_2$ abundances than those of G stars. However, this could also be caused by JWST target selection-induced degeneracy as cool and low-mass planets orbit cool stars in the 8-planet sample. 

Surface gravity determines the atmospheric scale height, which directly affects the length scales of atmospheric vertical mixing (K$_{zz}$) \citep{smith_estimation_1998}. High K$_{zz}$ could lead to depletion of CH$_4$ in the observable region of the atmosphere if the quench level is in the CO-dominated temperature-pressure region \citep{fortney_beyond_2020, fu_strong_2022}. Since CH$_4$ absorbs around 3.3 $\mu$m, lower CH$_4$ abundance would lead to a lower L band value and therefore higher SO$_2$-L index. High K$_{zz}$ could also change the H$_2$S and SO$_2$ abundances in the upper atmosphere by lifting up more H$_2$S and spreading out SO$_2$ to wider pressure levels. Changing surface gravity can also change the atmospheric TP profile and how deep the UV flux can penetrate. As temperature increases faster with pressure and atmospheric density is overall higher under lower surface gravity, the photochemical region moves to lower-pressure regions. We do not observe a strong linear correlation between surface gravity and SO$_2$-L.

\subsubsection{CO$_2$-L trends}

Similar to SO$_2$-L, we plotted CO$_2$-L versus mass, temperature, T$_{eff}$ and logg (Figure \ref{fig3}, Table \ref{tab:table2}). The CO$_2$ abundance is also expected to have a temperature dependency and it is largely not sensitive to disequilibrium chemistry within the sample temperature range \citep{moses_disequilibrium_2011}. To demonstrate this, we generated a set of forward models using \texttt{PLATON} \citep{zhang_platon_2020} assuming cloud-free, equilibrium chemistry, and isothermal TP profiles. We adopted seven metallicities (1x, 3x, 10x, 30x, 100x, 300x, and 1000x solar) and six absolute C/O ratios (0.1, 0.3 0.5, 0.7, 0.9, and 1.1) for the generic models. All models are generated under 1 Jupiter radius and 1 Jupiter mass planet. The exact planet mass and radius choices are not relevant as all models are then normalized by their respective scale heights. Next, CO$_2$-L is calculated using the same method as before. For the C/O=0.3 set of models, the CO$_2$-L index peaks around $\sim$900K for Z=30x solar models due to the prominent 4.3 $\mu$m CO$_2$ feature. The index values then decrease for models with lower Z, hotter or cooler temperatures. For the C/O=0.7 set of models, the CO$_2$-L index peaks at higher temperatures because of the larger 3.3 $\mu$m CH$_4$ feature driven by the higher C/O ratio, which drives up the L-band value and decreases the CO$_2$-L index.  
Solely based on the observed CO$_2$-L values, all planets are enhanced in metallicity beyond solar. However, the goal of these model tracks is not to precisely determine the metallicity and C/O for each planet, but to show the structures we expect in the diagram from equilibrium chemistry and isothermal assumptions, which are usually the starting points for more complex modeling efforts. 

\begin{table*}[]
\centering
\begin{tabular}{ccccccccccc}
\hline
\hline
Planet	&	T$_{eq}$ (K)	&	Mass (M$_J$)	&   logg (cgs)  &	T$_{eff}$ (K)	&   Z fraction  &	SO$_2$-L			&	CO$_2$-L			&	CO-L		\\
\hline
GJ 3470b	&	615$\pm$16	&	0.044$\pm$	0.005	&	2.82	$\pm$	0.06	&	3552$\pm$157	&	0.84	$\pm$	0.05	&	0.32	$\pm$	0.12	&	0.94	$\pm$	0.12	&	0.15	$\pm$	0.10	\\
HIP 67522b	&	1176$\pm$22	&	0.047$\pm$	0.016	&	2.19	$\pm$	0.15	&	5675$\pm$75	&	0.7	$\pm$	0.06	&	-0.71	$\pm$	0.15	&	3.80	$\pm$	0.18	&	1.04	$\pm$	0.14	\\
WASP 107b	&	770$\pm$60	&	0.1$\pm$	0.01	&	2.45	$\pm$	0.05	&	4425$\pm$70	&	0.18	$\pm$	0.03	&	0.15	$\pm$	0.14	&	4.70	$\pm$	0.16	&	0.48	$\pm$	0.13	\\
WASP 127b	&	1420$\pm$24	&	0.165$\pm$	0.02	&	2.38	$\pm$	0.06	&	5620$\pm$85	&	0.5	$\pm$	0.033	&	-1.55	$\pm$	0.28	&	2.64	$\pm$	0.39	&	-0.41	$\pm$	0.28	\\
WASP 39b	&	1166$\pm$14	&	0.28$\pm$	0.03	&	2.63	$\pm$	0.05	&	5485$\pm$50	&	0.19	$\pm$	0.04	&	-0.39	$\pm$	0.11	&	2.42	$\pm$	0.11	&	0.51	$\pm$	0.10	\\
WASP 39b	&	-	&	-		&	-			&		&	-			&	-0.08	$\pm$	0.11	&	2.89	$\pm$	0.14	&	0.32	$\pm$	0.12	\\
HD 209458b	&	1459$\pm$12	&	0.73$\pm$	0.04	&	2.97	$\pm$	0.03	&	6091$\pm$10	&	0.16	$\pm$	0.02	&	-0.82	$\pm$	0.06	&	0.46	$\pm$	0.09	&	-0.12	$\pm$	0.07	\\
HD 189733b	&	1209$\pm$11	&	1.13$\pm$	0.08	&	3.34	$\pm$	0.03	&	5052$\pm$16	&	0.13	$\pm$	0.03	&	-0.90	$\pm$	0.09	&	1.74	$\pm$	0.09	&	0.43	$\pm$	0.08	\\
WASP 121b	&	2450$\pm$8	&	1.157$\pm$	0.07	&	2.97	$\pm$	0.03	&	6776$\pm$138	&	0.13	$\pm$	0.02	&	-0.57	$\pm$	0.17	&	0.93	$\pm$	0.20	&	1.98	$\pm$	0.16	\\
\hline
\end{tabular}
\footnotesize{$^a$ NIRSpec PRISM spectrum from \citetalias{carter_benchmark_2024} \\ $^b$ NIRSpec G395H spectrum from \citetalias{carter_benchmark_2024}}\\
\caption{Planet parameters and corresponding SO$_2$-L, CO$_2$-L and CO-L index values. For WASP 39b, the PRISM index value is used instead of G395H. Our overall results are not sensitive to the choice of instrument. The potential causes for the difference between the two could be instrumental or astrophysical. Without knowing the ground truth spectrum, we can not determine which instrument has the more accurate spectrum. If any future calibration program demonstrates that G395H is more accurate than PRISM, the G395H index values from this table should be used instead.}
\label{tab:table2}
\end{table*}

\subsubsection{CO-L trends}

Unlike CO$_2$, CO features only vary weakly with changing abundance due to the strong triple bond. Therefore it is not a great metallicity tracer as varying CO abundance in the atmosphere does not translate into a prominent change in the CO spectral feature. However, since most of the carbon and oxygen ($>$99$\%$) are expected to be stored in CO and H$_2$O in hot ($>$1000K) H$_2$-dominated atmospheres \citep{moses_chemical_2013}, CO-L should trace the atmospheric bulk oxygen and carbon inventory for these hot Jupiters. We generated simple forward models with \texttt{PLATON} \citep{zhang_platon_2020} assuming equilibrium chemistry and isothermal TP profiles. At a given temperature, metallicity has a limited effect until 30x solar where the CO$_2$ becomes very prominent and lifts the CO band. On the other hand, C/O ratios have more significant effects at $>$1000K. As CH$_4$ are dissociated, C/O directly determines the relative CO to H$_2$O feature strength. At the ultra-hot Jupiter region (T$_{eq}>$2000K) where H$_2$O also starts to thermally dissociate \citep{arcangeli_h-_2018, fu_hubble76_2021, lothringer_extremely_2018}, CO-L increases further as CO becomes the main molecule to be present in the atmosphere.

We see a prominent CO feature in the ultra-hot Jupiter WASP-121b which is reflected in its high CO-L value. Temperature (r=0.46$\pm$0.07) and Teff (r=0.43$\pm$0.06) show moderate correlation with CO-L while mass (r=0.22$\pm$0.06) and logg (r=0.08$\pm$0.07) do not (Figure \ref{fig3_CO}). The correlations with temperature and Teff are mostly driven by WASP-121b which is the hottest planet which also orbits the hottest host star. Excluding WASP-121b, the correlation with temperature (-0.21$\pm$0.1) and Teff (-0.07$\pm$0.1) dropped significantly. The diverging model predictions on CO-L for hot to ultra-hot Jupiters make this a useful index to study population-level atmospheric C/O trends as more future data points come in.

\subsubsection{Population-level atmospheric metallicity enhancement}

Carbon dioxide is a powerful atmospheric metallicity indicator as its 4.3$\mu$m feature being sensitive to atmospheric metallicity changes \citep{lodders_atmospheric_2002, moses_disequilibrium_2011, moses_chemical_2013}. The abundance of CO$_2$ is also insensitive to disequilibrium processes such as vertical mixing and photochemistry \citep{moses_disequilibrium_2011}. On the other hand, sulfur dioxide is the product of disequilibrium chemistry processes, but it is also predicted to be a sensitive metallicity tracer. After accounting for the temperature-dependency of CO$_2$ and SO$_2$ abundances under minimal model assumptions as discussed above, we can use them to study population-level atmospheric metallicity enhancement and trends.

We interpolated the \texttt{PLATON} model grid (as discussed earlier) over metallicity, temperature, and C/O ratio parameter spaces. Metallicity values are then converted into planet mass based on seven different assumptions including seven uniform metallicities (1x, 3x, 10x, 30x, 100x, 300x, and 100x solar) and two mass-metallicity correlations (Solar system trend and freely fitted). We then measured the BIC (BIC = kln(n) + $\chi^2$) for each grid point relative to our sample (n=8) at six different fixed pollution-level C/O ratio values (0.1, 0.3, 0.5, 0.7, 0.9, and 1.1). Under these assumptions, freely fitted mass-metallicity correlation (k=3) provides the best fit (BIC=232) compared to that of uniform mass-metallicity (k=2, BIC=295) and Solar-system mass-metallicity trend (k=1, BIC=422). At the population level, low C/O ratios ($<$0.7) and super solar metallicity ($>$3x) are favored under our model assumptions (Figure \ref{fig_SS}, \ref{fig_BIC}).

Since we are only using a very limited number of free-fitting parameters (C/O ratio, mass-metallicity correlation) for the entire 8-planet sample, we do not expect to achieve the level of best-fit relative to the traditional retrieval approach which can use $>$10 free parameters per planet. Our restrictive model choice represents the foundation for future efforts to include more model complexity such as vertical mixing, photochemistry, C/O ratio variation, etc.

\subsection{SO$_2$-L versus CO$_2$-L}

From galaxies to stars to brown dwarfs, color-color diagrams have been used to describe different populations and evolutionary tracks. Here we introduce the color-color diagram for transiting exoplanets with SO$_2$-L versus CO$_2$-L (Figure \ref{fig4}). The goal is to identify any cluster or track of planet colors. For example, the upper left part of Figure \ref{fig4} is currently empty, indicating large CO$_2$ features may be related to SO$_2$ features. The emptiness of the lower left corner and the lack of low SO2-L planets could be due to the cloud continuum around 4$\mu$m. We cannot yet make definitive claims on the presence and cause of any patterns within the 8-planet sample due to the limited size and degeneracies between mass and temperature, e.g. the lower-mass planets are also cooler, orbiting later-type stars and possessing lower surface gravity. More planets are needed to determine if any empirical patterns emerge in the CO2-L versus SO2-L parameter space. 

Fortunately, more than 50 planets (M$_p>$0.03M$_J$) with 3-5 $\mu$m coverage have been or will be observed with JWST in cycles 1, 2, and 3 in transmission spectroscopy (Figure \ref{fig5}). If mass and temperature remain the main drivers that shape the transmission spectra of H$_2$-dominated atmospheres, the correlations between indices versus mass and temperature will remain strong. On the other hand, if we observe stronger scatters, that could point to two possible explanations: (1) the presence of other drivers such as host star SED, planet surface gravity, or other unknown parameters. (2) the planet sample is still insufficient and does not span wide enough parameter space.

\section{Conclusion}

We present a new empirical framework to characterize JWST infrared transmission spectra. Using four bands we aim to capture the spectral features that include all major carbon, nitrogen, oxygen, and sulfur-bearing molecules. We then correlated their relative values in scale heights with four main physical parameters to search for possible trends. We detect strong linear correlations between SO$_2$-L versus equilibrium temperature, indicating the presence of SO$_2$ is sensitive to temperature. Among the planet sample with T$_{eq}<$1400K, we also observe a correlation between SO$_2$-L versus planet mass and bulk metal fraction. This is consistent with SO$_2$ enhancement from increased metallicity \citep{tsai_photochemically_2023} with lower mass planets being more metal enriched \citep{welbanks_atmospheric_2022, thorngren_massmetallicity_2016}. We further explore these trends with a generic forward model grid to show how changing temperature and metallicity affects the SO$_2$-L values and better explain the observed trends. We also investigated CO$_2$-L and CO-L trends and they both vary significantly with temperature as predicted with simple equilibrium chemistry models. 

For H$_2$-dominated atmospheres, CO$_2$-L and SO$_2$ are expected to be sensitive to metallicity. By fitting the observed CO$_2$-L and SO$_2$-L values in the 8 spectra to our models under minimal population-level assumptions of uniform C/O ratio, equilibrium chemistry, and isothermal thermal structure, we find that short-period exoplanets are, in general, metal metal-enhanced, and the existence of a mass-metallicity relation is favored over uniform metallicity.

We are currently in the early stages of understanding exoplanet atmospheres on a population level compared to stars and galaxies. Although we have only focused on studying how the four main physical parameters (mass, temperature, host star, and surface gravity) drive observed transit spectra, other factors such as eccentricity, tidal heating, age, obliquity, etc. could all play a role in shaping the atmospheric composition and lead to additional scatter and deviation from the chemistry-focused atmospheric model predictions. If significant, these effects will show up as new features in the exoplanet palette diagrams with increasing future JWST atmospheric transit spectra ($\sim10^2$ planets). All of the existing or tentative correlations reported in this paper will also be directly tested with new JWST data in the next few years. \\

\noindent Contribution from S.P. on this project has been carried out within the framework of the National Centre of Competence in Research Planets supported by the Swiss National Science Foundation under grant 51NF40\_205606. S.P. acknowledges the financial support of the SNSF.

\clearpage

\bibliography{references}

\end{document}